\documentstyle[pre,preprint,aps,epsf]{revtex}
\draft
\tighten

\begin{document}
\author{Yuriy E. Kuzovlev}
\address{Donetsk Physics-Technical Institute of NASU, Donetsk, Ukraine}
\title{Exact Langevin equation for electrons in thermostat, 
Fokker-Planck kinetics and flicker noise}
\date{February 2001}
\maketitle

\begin{abstract}
Generalized Langevin equation for characteristic functional of many-electron
system dynamically interacting with a thermostat and besides subjected to
external perturbation and observation is derived and formulated in terms of
one-particle but stochastic density matrix. The scheme naturally includes
also direct electron-electron interactions. Corresponding approximate
Fokker-Plank kinetic equation is discussed. We point out that it possesses 
continuous degeneracy of stationary solution and therefore can hide 
long-correlated fluctuations of electron transport even if thermostat 
noise is short-correlated.
\end{abstract}

\subsection{Introduction}

The low-frequency flicker noise (1/f noise) is of growing interest among
scientists. Usually, investigators interpret it as manifestation of some
slow kinetics, for instance, of thermally activated switchings with a broad
variety of memory-times (life-times, relaxation times, etc.). From this
point of view, 1/f-noise says that present events in a system depend on its
old history. Opposite principal possibility is that 1/f-noise is due to
forgetting history, concretely, forgetting how many events took place in the
past, which means that a system has no reasons to keep well certain "number
of events per unit time". Then 1/f spectrum reflects only statistical
freedom of flow of events but not real cause-consequence connections between
them. From this point of view, 1/f noise in electronics and, say, 1/f
fluctuations of flowing of the river Niel are similar phenomena. General
ideology of this approach and its concrete realizations were published
elsewhere [1-11].

The two different ideas imply different programs for future theory. In our
approach, one should reconsider derivations of kinetic models of dissipation
(irreversibility) and noise from underlying Hamiltonian dynamics, avoiding
the ansatz that "numbers of events per unit time" (i.e. collision integrals,
etc.) have quite definite (and well achievable) limits as being averaged
over time. The suspicion that such quantities may be uncertain was claimed
and argued by Krylov [12] although with no relation to 1/f-noise. It must 
be emphasized that violation of ergodicity with respect to events, i.e.
transitions between phase space regions, does not mean violation of
ergodicity with respect to visiting these regions, therefore, usual ergodic
theorems are not under doubts. Also we have no doubts of importance of slow
kinetics (thermally activated or other).

In particular, it is interesting to elucidate whether many-electron system
interacting with a thermostat can possess 1/f type or other flicker
fluctuations in electron transport if autonomous thermostat dynamics itself
is free of such fluctuations. Any definite answer for this question would 
be equally useful. In this work we make several steps in this direction 
and argue that the answer is perhaply positive.

The object under our interest is characteristic functional of electron
current. First, we derive formally exact "Langevin equation" for fermions
subjected to Hamiltonian interaction with a thermostat, to external
dynamical perturbation (bias electric field) and besides to external
observation. Secondly, we show that operator-valued thermostat variables 
in this equation can be effectively replaced by doubled $c$-number-valued
variables ("noises"). As the consequence, the equation allows to perform
evaluation of fermionic statistical sums before averaging with respect to
thermostat. Therefore, the problem reduces to analysing statistical
properties of one-electron but instead stochastic ("twice randomized")
density matrix. Next, supposing that electron-thermostat interaction is 
weak and short-correlated, we derive approximate Fokker-Planck equation 
for probability distribution of the density matrix, and then consider
corresponding kinetic equation for its eigenvalues which just determine the
characteristic functional. Finally, we attract attention to that stationary
solution of the Fokker-Planck equation is continuously degenerated, and that
this fact can signify existence of long-correlated fluctuations in electron
transport (flicker type excess current noise). Mathematically rigorous 
proof (or refutation) of this statement needs in analysis of spectra of
complicated non-self-adjoint second-order differential operators. This 
will be next stage of the work.

\subsection{Model}

Let a quantum system consists of an essential ''dynamical subsystem'', with
Hamiltonian $H_e(t)$ , and a ''thermal bath'' (''thermostat''), with
Hamiltonian $H_b$ . We assume that (i) quantum states of these two parts
belong to different independent Hilbert spaces, and (ii) their interaction
is described by Hamiltonian 
\begin{equation}
H_{int}=\sum_jB_j*D_j
\end{equation}
where $D_j$ and $B_j$ act in dynamical space and bath space, respectively.
This is the case in most of physically meaningful models. For concreteness,
we may take in mind that dynamical subsystem (to be abbreviated ''DS'') is
formed by electrons obeying secondary quantization, while bath by bosons
(for instance, phonons). Thus the total Hamiltonian is 
\begin{equation}
H(t)=\,H_e(t)+H_b+H_{int}\,
\end{equation}

Besides the interaction, DS may be subjected to an external quasi-classical 
($c$-number) perturbation, $u(t)$ . In many cases it also can be 
described by bilinear form,
\begin{equation}
H_e(t)=H_e-u(t)Q\,\,,
\end{equation}
with $H_e$ being unperturbed Hamiltonian and $Q$ a conjugated quantum
variable. For example, if $u(t)$ is an external voltage then $Q$ represents
amount of charge transported by free (conducting) electrons, or if $u(t)$ 
is applied electric field then $Q$ is polarization. Generally speaking,
operator $Q$ does not have definite expression, but the corresponding flow
(current), $i[H(t),Q]$ , is well defined (square brackets will serve for
operator commutation, $[X,Y]=$ $XY-YX$ ). We assume that (iii) $Q$ belongs
to the dynamical subspace (or to some its extension) and (iv) $Q$ commutes
with all the operators $D_j$ , and thus $[H_{int},Q]$ $=0$ . This means that
thermal bath prodicing dissipation and decoherence in DS at the same time 
do not contribute to the current in a direct way, as usually in models of
electron transport. Then the current operator
\begin{equation}
J=i[H(t),Q]\,=i[H_e,Q]
\end{equation}
also purely belongs to the DS subspace, and 
\begin{equation}
\lbrack H_b,J]=0
\end{equation}
Of course, in any non-trivial case operators $D_j$ do not commute with free
DS Hamiltonian, $[H_e,D_j]\neq 0$ , and corespondingly $[J,D_j]\neq 0$ ,
although commutation $[H_e,J]=0$ is possible.

\subsection{Quantum characteristic functional}

The things under our interest will be fluctuations in DS, firstly in the
current, $J$ , and its response to external perturbation. To find these, one
should consider time-dependent operators in the Heisenberg representation,
\begin{equation}
J(t)=S^{+}(t)JS(t)\,,\,\,\,S(t)\equiv \overleftarrow{\exp }\left[
-i\int_0^tH(t^{\prime })dt^{\prime }\right] \,,
\end{equation}
where superscript ''+'' denotes Hermitian conjugation, and 
symbols $\overleftarrow{\exp }$ and $\overrightarrow{\exp }$ 
serve for chronological and anti-chronological time-ordering exponents, 
respectively. In the Heisenberg picture, fluctuations are described by 
statistical moments $\left\langle J(t_1)...J(t_n)\right\rangle $ , 
where angle brackets is averaging via some statistical ensemble of 
initial states, 
\begin{equation}
\left\langle ...\right\rangle \equiv \, \text{Tr }\,
\{...\}\rho _{in}\,\,,
\end{equation}
with $\rho _{in}$ being an initial density matrix.

To have real-valued moments, it is necessary to choose a certain
symmetrization rule, in other words, some quantum generalization of
classical characteristic functional 
\[
\Xi (v)=\left\langle \exp \left[ \int_0^tv(t^{\prime })J(t^{\prime
})dt^{\prime }\right] \right\rangle \,, 
\]
where $v(t)$ is arbitrary test function. The first and second
moments are indifferent to the choice. One of possible definitions is 
\begin{equation}
\Xi (v)\equiv \left\langle \overrightarrow{\exp }\left[ 
\frac 12 \int_0^tv(t^{\prime })J(t^{\prime })dt^{\prime }\right] 
\overleftarrow{\exp } \left[ 
\frac 12\int_0^tv(t^{\prime })J(t^{\prime })dt^{\prime }\right]
\right\rangle
\end{equation}
It corresponds to following time-ordered symmetric product: 
\[
J(t_1)...J(t_n)\equiv \{..\{J(t_1)\circ J(t_2)\}..\circ J(t_{n-1})\}\circ
J(t_n)\,\,\,\,\,\text{if }\,\,\, t_1>t_2>..>t_n\,, 
\]
where the circle means Jordan symmetric 
product, $X\circ Y\equiv $ $\frac 12 (XY+YX)$ .

The convenience of this variant of symmetrization is that it allows to
reduce calculation of the integral quantity $\Xi (v)$ to solution of a
differential equation. With the help of well known general recipes for
disentangling operator exponents we can rewrite expression (8) as
\begin{equation}
\Xi (v)=\left\langle \overrightarrow{\exp } 
\left( \int_0^t\left[ \frac 12
v(t^{\prime })J+iH(t^{\prime }) 
\right] dt^{\prime }\right) \overleftarrow{
\exp } \left( \int_0^t\left[ 
\frac 12v(t^{\prime })J-iH(t^{\prime }) \right] dt^{\prime }\right) 
\right\rangle
\end{equation}
Because of general properties of the Trace operation this means that 
\[
\Xi (v)=\text{Tr }\rho (t)\,, 
\]
where operator $\rho (t)$ obeys the equation 
\begin{equation}
\frac d{dt}\rho (t)=v(t)J\circ 
\rho (t)+i[\rho (t),H(t)] \equiv v(t)J\circ
\rho (t)+L(t)\rho (t)\,,
\end{equation}
with initial condition $\rho (0)=\rho _{in}$ . Equation (10) remains valid
also in classical limit, if replace $L(t)$ by Liouville operator and $J$ by
phase function of canonical variables, respectively. At $v(t)=0$ it turns
into mere von Neumann (Liouville) evolution equation for density matrix
(probability distribution). Hence, the testing is equivalent to
non-Hamiltonian perturbation which can not be introduced by means of Poisson
brackets.

\subsection{Free bath representation}

We will suppose that the only significant role of thermal bath is destroying
coherence of DS evolution, i.e. production of noise, dephasing and
dissipation. Otherwise (for instance, if the bath changes ground state of
electronic subsystem), we must redefine division of the whole system into
two parts. Then, it is reasonable to speak the free bath language in which
bath behaves independently on DS, while the latter is disturbed by bath and
governed by the Hamiltonian 
\begin{equation}
H(B,t)\equiv H_e(t)+\sum_jB_j(t)D_j
\end{equation}
Here the operators 
\[
B_j(t)\equiv \exp (iH_bt)B_j\exp (-iH_bt) 
\]
describe free bath evolution. Taking into account the equality (5), it is
easy to verify that expressions (6) and (9) are indeed identical to 
\begin{equation}
J(t)=S_B^{+}(t)JS_B(t)\,,\,\,S_B(t)\equiv \overleftarrow{\exp }\left[
-i\int_0^tH(B,t^{\prime })dt^{\prime }\right] \,,
\end{equation}
\begin{equation}
\Xi (v)=\left\langle \overrightarrow{\exp } 
\left( \int_0^t\left[ \frac 12
v(t^{\prime })J+iH(B,t^{\prime }) 
\right] dt^{\prime }\right) \overleftarrow{
\exp } \left( \int_0^t\left[ \frac 12v(t^{\prime })J-iH(B,t^{\prime })
\right] dt^{\prime }\right) \right\rangle
\end{equation}

\subsection{Quantum thermostat to classical noise conversion}

The ''free bath representation'' (12)-(13) allows to replace quantum 
bath by an effective classical disturbation (noise), or to be precise, by
pseudo-classical one. In perfect sense this can be made under assumption
that (v) initial statistical ensemble has the factored form 
\begin{equation}
\rho _{in}=\rho _e*\rho _b\,\,,
\end{equation}
with no initial correlations between two subsystems, and therefore the
complete averaging exactly divides into two stages: 
\begin{equation}
\left\langle ...\right\rangle =\left\langle 
\left\langle ... \right\rangle
_e\right\rangle _b\,\,,\,\, 
\left\langle ...\right\rangle _e\equiv 
\text{Tr}_e\,\{...\}\, \rho _e\,\,,\,
\,\,\left\langle ...\right \rangle _b\equiv 
\text{ Tr}_b\,\{...\}\,\rho _b\,,
\end{equation}
where Tr$_e$ and Tr$_b$ are traces in dynamical and bath subspaces.

Consider the construction 
\begin{equation}
\Xi (v)=\left\langle \overrightarrow{\exp } 
\left( \int_0^t\left[ \frac 12 v(t^{\prime })J+iH(\eta ,t^{\prime }) 
\right] dt^{\prime }\right) 
\overleftarrow{\exp }\left( 
\int_0^t\left[ \frac 12v(t^{\prime })J-iH(\xi ,t^{\prime }) 
\right] dt^{\prime }\right) \right\rangle
\end{equation}
with different left-hand and right-hand Hamiltonians defined 
by replacing $B(t)$ in (11) by $\eta (t)$ and $\xi (t)$ , 
and $\eta _j(t)$ and $\xi _j(t)$ being formally treated 
as classical ($c$-number valued) noises. Obviously,
this expression coinsides with (13) if 
interpret $\left\langle ...\right\rangle _b$ as 
averaging with respect to noises $\eta (t)$ and $\xi (t)$ 
under an appropriate choice for their statistics. To get the exact
identity, we should introduce their characteristic functional by 
\begin{equation}
\left\langle \exp \,\int \sum_j[a_j(t)\xi _j(t)+b_j(t) 
\eta _j(t)]dt\right\rangle _b\equiv
\end{equation}
\[
\equiv \text{Tr}_b\,\,\overrightarrow{\exp }\left[ \int
\sum_jb_j(t)B_j(t)dt\right] \overleftarrow{\exp }\left[ \int
\sum_ja_j(t)B_j(t)dt\right] \,\rho _b\,\,, 
\]
where $a_j(t)$ and $b_j(t)$ are test functions. As consequence, 
instead of (10), we can write 
\[
\Xi (v)=\left\langle \text{Tr}_e\,\rho (t)\right\rangle _b \,\,, 
\]
where now ''density matrix'' $\rho (t)$ obeys 
the generalized von Neumann equation 
\begin{equation}
\frac d{dt}\rho (t)=v(t)J\circ 
\rho (t)+L(t)\rho (t)\,\,, \,\,
\,L(t)\rho =i\{\rho H(\eta ,t)-H(\xi ,t)\rho \}\,,
\end{equation}
with initial condition $\rho (0)=\rho _e$ . It concerns DS only 
and evolves purely in the dynamical subspace, 
while $\eta (t)$ and $\xi (t)$ play a role
of noise perturbation externally applied to DS.

The ansatz (14) is natural approximation if the interaction $H_{int}$ is
sufficiently weak. But in general case we also may start from the ensemble
(14), if relating $\Xi (v)$ to later times than characteristic time of
mutual relaxation of subsystems, and thus considering their thermodynamical
interaction as result of dynamical time evolution.

\subsection{Generalized Langevin equation}

The convenience of the pseudo-classic representation introduced by Eqs.16-18
is that one can deal with $\eta (t)$ and $\xi (t)$ as objects which commute
with both themselves and arbitrary other objects. The pay for this
simplification is that DS evolution as determined by Eqs.16 and 18 becomes
non-unitary, because of $\eta (t)\neq \xi (t)$ . It is useful to introduce
new noises,
\[
\xi \equiv x+iy/2\,,\,\,\eta \equiv x-iy/2\,,
\,\,\,x=(\eta +\xi )/2\,,\,\,\,y=i(\eta -\xi ) 
\]
In their terms Eqs.17 and 18 read 
\begin{equation}
\left\langle \exp \,\int 
\sum_j[g_j(t)x_j(t)+f_j(t)y_j(t)]dt\right\rangle _b\equiv
\end{equation}
\[
\equiv \text{Tr}_b\,\,
\overrightarrow{\exp }\left\{ 
\int \sum_j\left[ \frac{ g_j(t)}2+if_j(t)\right] B_j(t)dt\right\} 
\overleftarrow{\exp }\left\{ \int
\sum_j\left[ \frac{g_j(t)}2-if_j(t)\right] B_j(t)dt\right\} 
\,\rho _b\, 
\]
with $g_j(t)$ and $f_j(t)$ being test functions, and Eq.18 takes the form
\begin{equation}
\frac d{dt}\rho (t)=v(t)J\circ \rho +L(t)\rho (t)\,,\,
\,\,L(t)\rho =\sum_jy_j(t)D_j\circ \rho +i[\rho ,H(x,t)]\,\,,
\end{equation}
where $H(x,t)$ is defined again by replacing $B(t)$ 
in (11), $H(x,t)=$ $H_e(t)+$ $\sum_jx_j(t)D_j$ . 
According to (19), in bath phase space $x(t)$
is associated with time-ordered symmetric multiplying, $B(t)\circ ...$ ,
while $y(t)$ with time-ordered commutation, $i[B(t),...]$ .

At $v(t)=0$ , Eq.20 (or equivalently Eq.18) turns into what can be named
generalized Langevin stochastic equation for DS interacting with a
thermostat. It differs from frequently used quantum Langevin equations by
the terms containing $y_j(t)$ . Clearly, just these terms destroy unitarity
of marginal DS evolution. Like the testing they represent non-Hamiltonian
perturbation of DS. Their physical meaning is the phase volume exchange
between DS and bath. Of course, the unitarity effectively restores after 
the complete averaging is performed. 

Notice that evidently dissipative terms appear in Eq.20 if divide 
products $x_j(t)\rho $ into the 
averages $\left\langle x_j(t)\rho \right\rangle _b$ 
and normalized forms $:x_j(t)\rho : $ . From the other hand, 
comparison of Eqs.10 and 20 shows that Tr$\,_e\,\rho (t)$ may be 
interpreted as mutual characteristic functional of
observables $J(t)$ and $D_j(t)\,$ as if governed by 
Hamiltonian $H(x,t)$ and tested
by $v(t)$ and $y_j(t)$ , respectively.

It should be emphasized that Eq.20 also directly extends to classical
mechanics if replace commutators by classical Poisson brackets. From the
point of view of classical intrinsic bath dynamics, multiplying by $y(t)$
looks as chronologically ordered action of the brackets,
\[
y(t)\Leftrightarrow \sum \frac{\partial B(t,q,p)}{\partial p}
\frac \partial {\partial q}-\sum 
\frac{\partial B(t,q,p)}{\partial q}\frac 
\partial {\partial p}\,\,, 
\]
where $q,p$ are bath canonical coordinates and 
momenta, $B(t,q,p)$ are $B(t)$
as expressed via initial values of $q,p$ , and 
differentiations $\partial /\partial q$ , $\partial /\partial p$ 
are oriented in the direction
of initial bath distribution $\rho _b=$ $\rho _b(q,p)$ .

\subsection{Effective bath noise statistics}

Thus $y_j(t)$ substitute superoperators even in classical case.
Nevertheless, even in quantum case we may treat 
them as well as $x(t)$ like $c$-number variables if 
only remember the definition (19) and carefully keep
in mind their time assigning. The evident profit 
of such a symbolic calculus is that now a 
''ready-made'' statistics of bath noises $x(t)$ and $y(t)$ may
serve as a model fit of the interaction. But underlying operator 
nature of $y(t)$ results in their unusual 
partly ''ghost'' effective statistical
properties. For instance, at $a(t)=0$ Eq.19 turns into
\[
\left\langle \exp \left\{ \int 
\sum_jb_j(t)y_j(t)dt\right\} \right\rangle _b=1\,\,, 
\]
that is all statistical moments of $y_j(t)$ are zero. 
In spite of this, $y_j(t)$ are non-zero, 
since they have non-zero cross correlations with $x_j(t)$ .

Of course, $x(t)$ and $y(t)$ are closely 
related being two manifestations of
the same intrinsic bath life. General connections between 
them follow from the properties of the ''trace'' procedure. 
In the stationary case, when $[H_b,\rho _b]$ $=0$ , 
for two-point correlators (which are translationally
invariant in this case as well as higher correlators) we have
\begin{equation}
K_{jm}(\tau )\equiv \text{Tr}_b\,\,B_j(\tau )B_m(0)\,\rho
_b\,=K_{mj}^{*}(-\tau )
\end{equation}
and correspondingly from Eqs.17 and 19 obtain 
\[
\left\langle \eta _j(\tau )\xi _m(0)
\right\rangle =K_{jm}(\tau )\,;\,\,
\left\langle \xi _j(\tau )\xi _m(0)\right\rangle =\left\langle 
\eta _j(\tau )\eta _m(0)\right\rangle ^{*}= 
\begin{tabular}{l}
$K_{jm}(\tau )\,,\,\,\tau >0$ \\ 
$K_{jm}^{*}(\tau )\,\,,$ $\tau <0$%
\end{tabular}
\]
\begin{equation}
K_{jm}^{(xx)}(\tau )\equiv 
\left\langle x_j(\tau )x_m(0)\right\rangle =\text{Re}
\,K_{jm}(\tau )\,;\,\,K_{jm}^{(xy)}(\tau )\equiv 
\left\langle x_j(\tau )y_m(0)\right\rangle = 
\begin{tabular}{l}
$2\,\,$Im\thinspace \thinspace $K_{jm}(\tau )\,,\,\,\tau >0$ \\ 
$0\,\,,\,\,\tau <0$%
\end{tabular}
\end{equation}
Here asterisk denotes complex conjugate, and subscript ''$b$'' is omitted.

Under the canonical bath distribution, 
\[
\rho _b\propto \exp (-H_b/\Theta )\,, 
\]
where $\Theta $ is bath temperature, additional relation takes place, 
\begin{equation}
K_{jm}(\tau -i/\Theta )=K_{mj}(-\tau )\,\,
\end{equation}
The latter if combined with (21) means that 
\[
K_{jm}^{(xx)}(\tau )=\int_0^\infty 
\cos (\omega \tau )\sigma _{jm}(\omega )d\omega \,\,,
\,\,K_{jm}^{(xy)}(\tau )=2\vartheta (\tau )\int_0^\infty 
\sin (\omega \tau )\tanh \left( \frac 
\omega {2\Theta }\right) \sigma _{jm}(\omega )d\omega 
\]
with some real non-negative spectral 
measure $\sigma _{jm}(\omega )$ 
and $\vartheta (\tau )$ being step function.

In the important specific case, when $B(t)$ represent a free bosonic field
and therefore self-commutator of $B(t)$ is $c$-number,
\[
i[B_j(t),B_m(t^{\prime })]=2\,\text{Im }K_{mj}(t^{\prime }-t)\,\,, 
\]
for any functional $\Phi (x,y)$ the Eq.19 implies 
\[
\left\langle y_j(t)\Phi (x,y)\right\rangle =
\left\langle \int_{t^{\prime }>t}\sum_m[2\,
\text{Im }K_{mj}(t^{\prime }-t)]
\frac{\delta \Phi (x,y)}{\delta x_m(t^{\prime })}dt^{\prime }
\right\rangle 
\]
In this sense, in case of free bosonic thermostat
\begin{equation}
y_j(t)=\sum_m \int_{t^{\prime }>t}dt^{\prime }[2\,
\text{Im }K_{mj}(t^{\prime }-t)]
\frac \delta {\delta x_m(t^{\prime })}
\end{equation}

\subsection{Ghost fields and dissipation}

According to relations (22) and (24), 
all ghost noises $y(t)$ are correlated
with more late $x(t)$ only. The meaning of this 
property becomes trivial if rewrite Eq.19 in the form
\[
\left\langle \exp \,\int 
\sum_j[g_j(t)x_j(t)+f_j(t)y_j(t)]dt\right\rangle _b=\,\, 
\]
\[
=\text{Tr}_b\,\,\overrightarrow{\exp }\left\{ \frac 12\int
\sum_jg_j(t)B_j(t,f)dt\right\} \overleftarrow{\exp }
\left\{ \frac 12\int
\sum_jg_j(t)B_j(t,f)dt\right\} \,\rho _b\,\,, 
\]
where $B_j(t,f)$ correspond to non-free bath evolution perturbed by
classical forces $f(t)$ ,
\[
B_j(t,f)=\overrightarrow{\exp }
\left[ i\int_0^tH_b(t^{\prime })dt^{\prime
}\right] B_j\overleftarrow{\exp }
\left[ -i\int_0^tH_b(t^{\prime })dt^{\prime
}\right] \,\,,\,\,\,H_b(t)\equiv H_b+\sum_jf_j(t)B_j 
\]
Differentiation of this identity by $g(t)$ and $f(t)$ shows that cross
correlators are nothing but bath response functions to the perturbation: 
\begin{equation}
\left\langle \prod_jx(t_j)\prod_my(t_m^{\prime })
\right\rangle =\left[
\prod_m\frac \delta {\delta f(t_m^{\prime })}\left\langle
\prod_jB(t_j,f)\right\rangle \right] _{f=0}
\end{equation}
Consequently, for all higher-order correlators
\begin{equation}
\left\langle \prod_jx(t_j)\prod_my(t_m^{\prime })
\right\rangle =0\,\,\,\,\text{if \thinspace max}
\{t_m^{\prime }\}>\text{max}\{t_j\}\,\,,
\end{equation}
in accordance with causality principle. This property just guarantees
eventual safety of the unitarity and of summary phase volume of two
subsystems. Due to (26), as one can see from Eq.20, at $v(t)=0$ 
always Tr$_e$  $\,\left\langle \rho (t)\right\rangle _b$  $=1$ .

Because of obvious dual symmetry of the subsystems, 
with reference to bath $f(t)$ and $g(t)$ play in Eq.25 
the same roles of exciting forces and test
functions, respectively, as $x(t)$ and $y(t)$ in Eq.20 with reference to DS.
It is not hard to guess that the intention of variables $y(t)$ in the
symbolic bath noise calculus is involving a back action of DS onto bath,
i.e. reverse dissipative energy flow which must ensure conservation of total
energy. If $y(t)$ were absent then only non-reciprocal pumping of DS by
forces $x(t)$ would be present.

\subsection{Fermionic dynamical subsystem}

Let DS is purely fermionic representing free conducting electrons (plus 
some electron depository if necessary). Our main ansatz will be that 
(vi) all the operators related to this subsystem
have the quadratic form
\begin{equation}
A=\sum_{\alpha \beta }A_{\alpha \beta }C_\alpha ^{+}C_\beta
\end{equation}
where $C^{+}$ ($C$) are fermionic creation (destruction) operators, 
indices $\alpha $, $\beta $ enumerate some complete 
set of one-fermion states, and $A$
stands for $H_e$ , $D_j$ , $Q$ and consequently $J$ . Thus the Hamiltonian
(2) does not contain direct electron-electron interactions like the
Coulombian one. But this simplification implies no principal losses, since
it is known that such interactions can be introduced by means of auxiliary
actual or ghost bosonic degrees of freedom included to both $H_b$ and the
set $B_j$ .

Indeed, let $H_e$ is quadratic and we add to it ''Coulombian'' or any other
pair interaction described by the Hamiltonian
\[
H_c=\sum_{\alpha \beta }V_{\alpha \beta }N_\alpha N_\beta
\,\,,\,\,\,N_\alpha \equiv C_\alpha ^{+}C_\alpha 
\]
(of course, here $\alpha $, $\beta $ may relate to different set of electron
states than in (27)). Then Eq.20 turns into
\[
\frac d{dt}\rho (t)=v(t)J\circ \rho (t)+L(t)\rho (t)+i[\rho (t),H_c] 
\]
It is easy verifiable that its solution can be represented in the form 
\begin{equation}
\rho (t)=\left\langle \widetilde{\rho }(t)\right\rangle \,\,,
\,\,\frac d{dt}\widetilde{\rho }(t)=v(t)J\circ 
\widetilde{\rho }(t)+L(t)\widetilde{\rho }(t)+L^{\prime }(t)
\widetilde{\rho }(t)\,\,,
\end{equation}
\[
L^{\prime }(t)\widetilde{\rho }\equiv 
\sum_\alpha y_\alpha ^{\prime }(t)N_\alpha \circ 
\widetilde{\rho }+i\sum_\alpha x_\alpha ^{\prime }(t)[
\widetilde{\rho },N_\alpha ]\,\,, 
\]
where $x^{\prime }(t)$ and $y^{\prime }(t)$ are 
Gaussian white noises with correlators
\begin{equation}
\left\langle x^{\prime }(t)x^{\prime }(t_0)\right\rangle =
\left\langle y^{\prime }(t)y^{\prime }(t_0)\right\rangle =0\,,
\,\,\left\langle x_\alpha ^{\prime }(t)y_\beta ^{\prime }(t_0)
\right\rangle =2V_{\alpha \beta }\delta (t-t_0)\,,
\end{equation}
and angle brackets in (28) notate averaging with respect to these white
noises if mentioned in the sense of Stratonovich (not of Ito).

This is merely the variant of so called Stratonovich transformation. Now all
fermionic operators in the evolution equation (28) become quadratic. Hence,
after including $x^{\prime }(t)$ and $y^{\prime }(t)$ to the set of bath
noises (with operators $N_\alpha $ in place of $D_j$ ) electron-electron
interaction reduces to the above general stochastic scheme. 
Because of $\left\langle x^{\prime }x^{\prime }\right\rangle =0$ , 
one may say that $x^{\prime }(t)$ and $y^{\prime }(t)$ 
represent interaction with zero-energy bath.

\subsection{Fermions to matrices conversion}

To complete our simplifications, in addition to (14) we assume (vii) an
initial distribution of the electron DS be canonically produced by
quadratic-type operators. For instance, 
\begin{equation}
\rho _e=\frac{\exp (-H_0/T)
\exp (\mu \sum_\alpha N_\alpha /T)}{\text{Tr}_e\{\,
\exp (-H_0/T)\exp (\mu \sum_\alpha N_\alpha /T)\}}\,\,,
\end{equation}
where $H_0$ is some quadratic operator, 
and sum $\sum_\alpha N_\alpha $ is
the operator of total number of electrons (it commutes with any other
operators). In particular case, $H_0$ may coinside with $H_e$ . 
If wanting to fix a number of electrons, $N_{el}$ , 
instead of chemical potential, $\mu $ , 
we may consider the composition of grand canonical distributions, 
\begin{equation}
\rho _e=\frac{\delta (N_{el}-\sum_\alpha N_\alpha )\,
\exp (-H_0/T)}{\text{Tr}_e\,\{\delta (N_{el}-\sum_\alpha N_\alpha )\,
\exp (-H_0/T)\}}\,\,,\,
\,\,\delta (n)=\int_{-\pi }^\pi e^{in\varphi }
\frac{d\varphi }{2\pi } \,,
\end{equation}
with variety of imaginary chemical ponentials, $i\varphi T$ . In this
approach, thermodynamical effects of Coulombian interaction, if included
into dynamics as prescripted by Eqs.28 and 29 (but with the averaging put
off to a later stage), will come into play after a time of evolution.

Then, clearly, in Eq.16 we can confine ourselves by  
calculation of only specific fermionic spurs Tr$_e$ $\{\prod_nM_n\}$ , 
where each multiplier is an exponent of one of three 
sorts, $\exp (A)\,\,$, 
  $\overleftarrow{\exp }\{\int A(t)dt\}\,$ , 
or $\overrightarrow{\exp }\{\int A(t)dt\}$ , 
with $A$ being quadratic fermionic operator in the sense of
(27). This calculation is easy performable if take into account three
remarkable facts, namely, (1) any exponent of the second or third sort can
be written as an exponent of the first sort, (2) product of two such
''quadratic exponents'' can be represented as a single quadratic exponent
and (3) there is one-to-one correspondence between such 
operator expressions and analogous matrix expressions. In summary, 
\begin{equation}
\exp (\sum X_{\alpha \beta }C_\alpha ^{+}C_\beta )
\exp (\sum Y_{\alpha \beta }C_\alpha ^{+}C_\beta )=
\exp (\sum Z_{\alpha \beta }C_\alpha ^{+}C_\beta )\,\,
\Leftrightarrow \,\,\,e^Xe^Y=e^Z
\end{equation}
\begin{equation}
\overleftarrow{\exp }\left[ \int 
\sum X_{\alpha \beta }(t)C_\alpha ^{+}C_\beta dt\right] =
\exp (\sum Z_{\alpha \beta }C_\alpha ^{+}C_\beta )\,\,\,
\Leftrightarrow \,\,\,
\overleftarrow{\exp }\left[ \int X(t)dt\right] =e^Z
\end{equation}
Here $X_{\alpha \beta }\equiv $ $(X)_{\alpha \beta }$ , and so on, and the
right-hand sides are meant in the sense of matrix functions. These relations
can be proved if use the commutator identity
\[
\lbrack \sum X_{\alpha \beta }C_\alpha ^{+}C_\beta \,,
\sum Y_{\alpha \beta }C_\alpha ^{+}C_\beta ]=
\sum [X,Y]_{\alpha \beta }C_\alpha ^{+}C_\beta 
\]

Hence, any product of quadratic fermionic exponents can be transformed into
a single such exponent, for which it is known that 
\begin{equation}
\text{Tr}_e\text{ }
\exp (\sum Z_{\alpha \beta }C_\alpha ^{+}C_\beta )=\det
\,(1+e^Z)\,=\exp \,\text{Sp}\,\ln (1+e^Z)
\end{equation}
with $\det \,(...)$ being matrix determinant, 
and  Sp $M$ $\equiv $ $\sum_\alpha M_{\alpha \alpha }$ for 
any matrix $M$ . After that, according
to (32) and (33), the exponent $e^Z$ may be back unfolded as the exact
matrix copy of initial operator product. Therefore, if denote quadratic
operators and their exponents by the same letters as their matrix copies
then
\begin{equation}
\text{Tr}_e\,\{\prod_nM_n\}=\det \,\{1+\prod_nM_n\}
\end{equation}
Although $x(t)$ and especially $y(t)$ are to some extent ghost variables, we
can pass them through all these transformations like scalar $c$-number
variables, since they by their definition commute with arbitrary objects and
are uniquely marked by their time arguments.

\subsection{Secondary randomization instead of secondary quantization}

Let $\rho _0$ be matrix copy of quadratic 
fermionic exponent $\exp (-H_0/T)$ , 
with $H_0$ now defined in the space of one-electron states. Then, as it
follows from previous section, in case of grand canonical ensemble (30) the
characteristic function (16) can be represented by 
\begin{equation}
\Xi (v)=\frac{\left\langle \det [1+z\rho (t)]
\right\rangle _b}{\det (1+z\rho _0)}=
\frac{\left\langle \exp \,\text{Sp}\,
\ln [1+z\rho (t)]\right\rangle _b}{\exp \,\text{Sp}\,
\ln (1+z\rho _0)}\,\,,\,\,z\equiv e^{\mu /T}\,\,,
\end{equation}
where time-dependent one-particle density matrix 
\begin{equation}
\rho (t)=\overleftarrow{\exp } \left( \int_0^t\left[ 
\frac 12v(t^{\prime })J-iH(\xi ,t^{\prime })\right] dt^{\prime }\right) 
\rho _0\,\overrightarrow{\exp }\left( \int_0^t\left[ 
\frac 12v(t^{\prime })J+iH(\eta ,t^{\prime })\right] dt^{\prime }\right)
\end{equation}
is matrix copy of the above considered $\rho (t)$ . It obeys exact matrix
analogies of Langevin equations (18) and (20), but now under initial
condition $\rho (0)=\rho _0$ . The case of canonical ensemble (31) can be
reduced to calculating the same quantities as in (36): 
\begin{equation}
\Xi (v)=\left[ \left( \frac \partial {\partial z}\right)
^{N_{el}}\left\langle \exp \, \text{Sp}\,\ln 
\{1+z\rho (t)\}\right\rangle _b\right] _{z=0}\left[ 
\left( \frac \partial {\partial z}\right) ^{N_{el}}\exp \,\text{Sp}\,
\ln (1+z\rho _0)\right] _{z=0}^{-1}\,
\end{equation}

Now, when Fermi statistics is explicitly took into account, we may pay
our attention to one-electron version of Langevin equations (18) and (20)
and analyse statistical properties of one-electron density matrix (37) (or
corresponding wave functions) as evolving under influence by random
processes $x(t)$ and $y(t)$ . If we first performed bath averaging 
we would deal with evolution of many secondary quantized fermions. 
Instead we will deal with twice random (quantumly and
excitedly) motion of single particle.

The above consideration clearly demonstrated that we have all rights to 
deal with $x(t)$ and $y(t)$ as real-valued variables and correspondingly 
with $\eta (t)$ and $\xi (t)$ as complex conjugated 
variables, $\eta (t)=$ $(\xi (t))^{*}$ . 
Hence, if suppose test function $v(t)$ be also real-valued then
we can formally treat $\rho (t)$ as Hermitian matrix and write 
\begin{equation}
\rho (t)=S(t,t_0)\rho (t_0)S^{+}(t,t_0)\,\,\,,\,\,S(t,t_0)\equiv 
\overleftarrow{\exp }\left( \int_{t_0}^t\left[ \frac 12v(t^{\prime
})J-iH(\xi ,t^{\prime })\right] dt^{\prime }\right)
\end{equation}
In general, $S(t,t_0)$ is not unitary transformation, $SS^{+}\neq 1$ .

\subsection{Grand canonical ensemble}

Consider general relations what follow from one-particle matrix versions of
Langevin equations (18) and (20). It is easy to see from Eq.20 that
\[
\frac d{dt}\,\text{Sp}\,F(\rho )=v(t)\,
\text{Sp}\,J\rho \frac{dF(\rho )}{d\rho }+\sum_jy_j(t)\,
\text{Sp}\,D_j\rho \frac{dF(\rho )}{d\rho } 
\]
for any function $F(\rho )=$ $F(\rho (t))$ . 
Therefore, if putting on $F(\rho )=$ $\ln (1+z\rho )$ , 
we may rewrite Eq.36 as 
\begin{equation}
\Xi (v)=\left\langle \exp \left[ \int v(t)\,
\text{Sp}\,Jf(t)\,dt+\sum_j\int y_j(t)\,
\text{Sp}\,D_jf(t)\,dt\right] \right\rangle _b\,,\,\,\,f(t)\equiv 
\frac{z\rho (t)}{1+z\rho (t)}\,
\end{equation}
Clearly, matrix $f(t)$ reflects occupancy of one-electron states in factual
many-electron system. According to Eq.20, $f(t)$ is solution to the
non-linear stochastic equation 
\begin{equation}
\frac{df}{dt}=\frac{v(t)}2[fJ(1-f)+(1-f)Jf]+
\sum_j\frac{y_j(t)}2 [fD_j(1-f)+(1-f)D_jf]+i[f,H(x,t)]
\end{equation}

To find statistical moments of the current 
we have to differentiate $\Xi (v)$
by test function $v(t)$ and then turn it into zero. But the terms
in (40) containing $y_j(t)$ remain and work even at $v(t)$ $=0$ .
Hence, it is comfortable to introduce special notation 
for bath averaging under $v(t)=0$ :
\[
\left\langle \Phi (x,y)\right\rangle _0\equiv 
\left\langle \left[ \Phi (x,y)\,\exp \,
\sum_j\int y_j(t)\,
\text{Sp}\,D_jf(t)\,dt\right] _{v=0}\right\rangle _b 
\]
with $\Phi (x,y)$ being arbitrary functional of bath influence. Due to the
property (26), $\left\langle 1\right\rangle _0=1$ . Because of (26), here as
well as in (40) in the exponent we may omit limits of the integration. Then
the mean current is 
\begin{equation}
\left\langle J(t)\right\rangle =\left[ \delta \Xi (v)/
\delta v(t)\right] _{v=0}=\left\langle 
\text{Sp}\,Jf(t)\right\rangle _0
\end{equation}
The current correlation function produced 
by second-order derivative of $\Xi(v)$ generally looks more complicated:
\begin{equation}
\left\langle J(t)J(t_0)\right\rangle =\left[ \delta ^2\Xi (v)/
\delta v(t)\delta v(t_0)\right] _{v=0}=
\end{equation}
\[
=\left\langle \text{Sp}\,\{f(t)J\}\,\,\text{Sp\thinspace }
\,S(t,t_0)[J\circ z\rho (t_0)]S^{+}(t,t_0)\{1+z\rho (t)\}^{-1}
\right\rangle _0+ 
\]
\[
+\left\langle \text{Sp}\,J\{1+z\rho (t)\}^{-1}\,S(t,t_0)[J
\circ z\rho (t_0)]S^{+}(t,t_0)\{1+z\rho (t)\}^{-1}\right\rangle _0\,, 
\]
where all matrices correspond to $v(t)$ $=0$ , 
and it is assumed that $t>t_0 $ .

\subsection{Non-degenerated electrons}

Let number of electrons is small as compared with total number of
one-electron states, $N_{el}<<N$ , $N\equiv $ Sp 1 , to the extent that all
mean occupancies are significantly smaller than unit. Then, in the frame of
canonical ensemble, Eq.38 approximately transforms into 
\begin{equation}
\Xi (v)\approx \left\langle [\text{Sp\thinspace }\,
\widetilde{\rho } (t)]^{N_{el}}\right\rangle _b\,\,,\,
\,\widetilde{\rho }(t)\equiv \frac{\,\rho (t)}{\text{Sp\thinspace }\,
\rho _0}
\end{equation}
Here the normalized density matrix $\widetilde{\rho }(t)$ , 
of course, also obeys Eqs.18 and 20. In its tems, 
\begin{equation}
\left\langle J(t)\right\rangle =N_{el}\left\langle \text{Sp\thinspace }
\,J \widetilde{\rho }(t)\right\rangle _{N_{el}-1}
\end{equation}
\begin{equation}
\left\langle J(t)J(t_0)\right\rangle =N_{el}(N_{el}-1)\left\langle 
\text{Sp} \,\{J\widetilde{\rho }(t)\}\,
\text{Sp}\,\{S(t,t_0)[J\circ 
\widetilde{\rho } (t_0)]S^{+}(t,t_0)\}\right\rangle _{N_{el}-2}+
\end{equation}
\[
+N_{el}\left\langle \text{Sp}\,JS(t,t_0)[J\circ 
\widetilde{\rho }(t_0)]S^{+}(t,t_0)\right\rangle _{N_{el}-1}\,\,, 
\]
where $t>t_0$ , and now the bath averaging at $v(t)$ $=0$ is defined by
\[
\left\langle \Phi (x,y)\right\rangle _n\equiv \left\langle 
\Phi (x,y)\,[\text{Sp\thinspace }\,
\widetilde{\rho }(t)]^n\right\rangle _b\approx
\left\langle \Phi (x,y)\,
\exp \,n\sum_j\int y_j(t)\,
\text{Sp}\,D_j\widetilde{\rho }(t)\,dt\right\rangle _b 
\]
(the latter expression relates to large $N$ and $n$ 
under small ratio $n/N$ $<<1$ ).

Since generally Sp $\,\widetilde{\rho }(t)$ $\neq 1$ , we see that
even in dilute electron gas contribution of any separate electron to total
current and its fluctuations statistically depends on all other electrons.
The matter is that all they feel the same thermostat and thus the latter
becomes mediator of electron-electron correlations if not interactions.
Since non-degenerated electrons obey Boltzmannian statistics, formulas
(44)-(46) relate also to classical limit.

\subsection{Diagonalization of stochastic density matrix}

According to Eqs.36, 38 and 44, in principle, to find the characteristic
functional we need in eigenvalues, $\lambda _k$ ($k=1..N$), of the
one-electron density matrix only. In view of Eq.39, 
we can treat $\rho (t)$ as Hermitian operator and 
write $\rho (t)=$ $\Psi (t)\lambda (t)\Psi ^{+}(t)$ , 
where $\lambda =$ diag$\{\lambda _k\}$ with real elements 
and $\Psi (t)$ is unitary transformation. 
Then in grand canonical ensemble 
\begin{equation}
\Xi (v)=\frac{\int \prod_{k=1}^N[1+z\lambda _k]W_\lambda (t,\lambda
)d\lambda }{\prod_{k=1}^N[1+z\lambda _{0k}]}\,\,\,,
\end{equation}
where $W_\lambda (t,\lambda )\equiv $ 
  $W_\lambda (t,\lambda _1,..,\lambda _N) $ 
is probability density distribution of the eigenvalues, 
  $d\lambda =$ $\prod_{k=1}^Nd\lambda _k$ , 
and $\lambda _{0k}$ are their initial values. In
ensemble with a fixed number of particles 
\begin{equation}
\Xi (v)=\int \left( \sum_{k_1<k_2<...}\lambda _{k_1}\lambda _{k_2}..
\lambda _{k_{N_{el}}}\right) W_\lambda (t,\lambda )d\lambda \left(
\sum_{k_1<k_2<...}\lambda _{0k_1}\lambda _{0k_2}..
\lambda _{0k_{N_{el}}}\right) ^{-1}\approx
\end{equation}
\[
\approx \int \left( \sum_{k=1}^N
\lambda _k\right) ^{N_{el}}W_\lambda (t,\lambda )d\lambda 
\left( \sum_{k=1}^N\lambda _{0k}\right) ^{-N_{el}}\,, 
\]
where the latter expression relates to dilute gas.

To realize this most adequate approach to the problem, we should combine
evolution equation (20) and the first-order perturbation theory to describe
infinitesimal time steps of the diagonalization. After standard
manipulations we obtain coupled evolution equations for eigenvalues and
eigenfunctions of density matrix: 
\begin{equation}
\frac{d\lambda _k}{dt}=\left[ v(t)(\Psi _k^{*},J\Psi _k)+
\sum_jy_j(t)(\Psi _k^{*},D_j\Psi _k)\right] \lambda _k
\end{equation}
\begin{equation}
\frac{d\Psi _k}{dt}=\frac 12\sum_{m\neq k}\Psi _m\frac{\lambda _k+
\lambda _m }{\lambda _k-\lambda _m}
\left[ v(t)(\Psi _m^{*},J\Psi _k)+\sum_jy_j(t)(
\Psi _m^{*},D_j\Psi _k)\right] -iH(x,t)\Psi _k
\end{equation}
Here $(a,b)=$ $\sum a_\alpha b_\alpha $ means pseudo-scalar product of two
complex vectors (thus $(a^{*},b)$ is usual scalar product) , and we suppose
that eigenvalues are non-degenerated. The latter assumption is not too
restrictive since $\lambda _k$ will acquire tendency to mutual repulsion.

\subsection{Weak noise and Fokker-Planck equation}

Let us emphasize that in fact the Langevin stochastic equations (49) and
(50), as well as Eqs.18 and 20, are still exact equations, in the sense that
they represent explicit mould of underlying detail Hamiltonian dynamics of
the electron-bath interaction. The natural next stage of the investigation
is derivation of corresponding kinetic equation for probability 
distribution of $\lambda $ and $\Psi $ (and thus for probability 
functional of $\rho (t)$ ).

If $x(t)$ and $y(t)$ were white noises the result would be time-local
Fokker-Planck (or Kolmogorov) equation again exactly reflecting the
Hamiltonian dynamics. As Eqs.28 and 29 demonstrate, this is just the case
when the noises represent time-local electron-electron interaction only. 
We put off this interesting example for the future. Another such case 
is $y(t)=0 $ while $x(t)$ being white noise which may be interpreted
as the infinite bath temperature limit. We will touch it below.

In general, exact kinetic equation is inevitably non-local since at any
finite bath temperature the relation (23) forbids to at once turn all the
correlators (22) into delta-functions. But we believe that a time-local
kinetic equation will be as usually good approximation if the bath noise is
sufficiently weak and short correlated. Moreover, under such assumptions we
can use simplest one-loop approximation resulting in a Fokker-Planck
equation. For any set of stochastic equations
\[
\frac{dZ_k}{dt}=A_k(t,Z)+\sum_\alpha x_\alpha (t)B_{\alpha k}(Z)+
\sum_\alpha y_\alpha (t)C_{\alpha k}(Z)\,\,, 
\]
with the same structure as Eqs.49 and 50, this approximation yields
\[
\frac{\partial W}{\partial t}=
\Lambda (t)W=\Lambda _f(t)W-\sum_k\frac 
\partial {\partial Z_k}F_k(t)\,W\,,\,
\,\,\Lambda _f(t)\equiv -\sum_k\frac 
\partial {\partial Z_k}A_k(t,Z)\,\,, 
\]
where $W=W(t,Z)$ is probability density function, operator $\Lambda _f(t)$
corresponds to free evolution and operators $F_k(t)$ describe the bath noise
contribution to the probability flow,
\[
F_k(t)=-\sum B_{\alpha k}(Z)\int_0^\infty 
\left\langle x_\alpha (\tau )x_\beta (0)\right\rangle e^{\tau 
\Lambda _f(t)}\frac \partial {\partial Z_m} B_{\beta m}(Z)e^{-\tau 
\Lambda _f(t)}d\tau - 
\]
\[
-\sum B_{\alpha k}(Z)\int_0^\infty 
\left\langle x_\alpha (\tau )y_\beta (0)
\right\rangle e^{\tau \Lambda _f(t)}\frac 
\partial {\partial Z_m}C_{\beta m}(Z)e^{-\tau \Lambda _f(t)}d\tau 
\]
(supposing noises are stationary and have zero mean values).

\subsection{Kinetics of probability measure of density matrix}

When applying this scheme it to our Eqs.49 and 50 we 
may neglect a contribution by $v(t)$
to the exponents like $\exp [\pm \tau \,\Lambda _f(t)]$ in the above
integrals, since under weak and short correlated bath noise
this contribution might result in relatively small corrections only
(physically, this means that we neglect displacements of an electron 
during its short collisions with bath excitations).

Then Fokker-Planck equation for the mutual probability density distribution
of the eigenvalues and eigenvectors, $W(t,\lambda ,\Psi )$ , looks as 
\begin{equation}
\frac{\partial W}{\partial t}=
\Lambda (t)W=[v(t)\Lambda _{test}+
\Lambda _{free}(t)+\Lambda _{xx}(t)+\Lambda _{xy}(t)]W\,\,,
\end{equation}
where kinetic operators $\Lambda _{test}$ , $\Lambda _{free}$ , 
  $\Lambda _{xx}$ and $\Lambda _{xy}$ describe testing of electron DS, 
its free evolution as if in absense of noise, 
its stochastic pumping by $x(t)$ which
itself influences $\Psi $ only, and common action of $y(t)$ and $x(t)$ which
leads to dissipation and besides to fluctuations of $\lambda _k$ ,
respectively. These four parts of $\Lambda (t)$ directly correspond to four
different type terms in stochastic equations (49)-(50). The first is 
\begin{equation}
\Lambda _{test}=-\sum_k\frac \partial {\partial 
\lambda _k}\lambda _k(\Psi _k^{*},J\Psi _k)-
\frac 12\sum_{k\neq m}\frac{\lambda _k+\lambda _m}{\lambda _k-
\lambda _m}\left[ \left( \frac 
\partial {\partial \Psi _k},\Psi _m\right) (\Psi _m^{*},J\Psi _k)+C.C.
\right]
\end{equation}
with $\partial /\partial \Psi _k$ being gradients 
and $(\partial /\partial \Psi _k)^{*}$ 
  $\equiv \partial /\partial \Psi _k^{*}$ .

To express other parts of the kinetic equation (51), for any Hermitian
matrix $M$ let us define first-order differential operator
\[
\Omega \{M\}\equiv i\sum_k\left[ \left( \frac \partial {\partial 
\Psi _k} ,M\Psi _k\right) -\left( \frac \partial {\partial 
\Psi _k^{*}},M^{*}\Psi _k^{*}\right) \right] 
\]
Obviously, $\Lambda _{free}(t)$ is such 
operator, $\Lambda _{free}(t)=$ $\Omega \{H_e(t)\}$ , 
while effect of noise $x(t)$ is described by the second-order operator 
\begin{equation}
\Lambda _{xx}(t)=\sum_{\alpha \beta }
\Omega \{D_\alpha \}\Omega \left\{
\int_0^\infty K_{\alpha \beta }^{(xx)}(\tau )
\exp [-iH_e(t)\tau ]D_\beta
\exp [iH_e(t)\tau ]d\tau \right\}
\end{equation}
The second-order operator $\Lambda _{xy}(t)$ contains $x$-$y$-correlators
instead of $x$-$x$ ones, 
\begin{equation}
\Lambda _{xy}(t)=-\Omega \{D_\alpha \}\frac 
\partial {\partial \lambda _k} 
\lambda _k\left( \Psi _k^{*},\int_0^\infty K_{\alpha \beta }^{(xy)}(
\tau )e^{-iH_e(t)\tau }D_\beta e^{iH_e(t)\tau }d\tau \Psi _k\right) -
\end{equation}
\[
-\frac 12\Omega \{D_\alpha \}\frac{\lambda _k+\lambda _m}{\lambda _k-
\lambda _m}\left[ \left( \frac 
\partial {\partial \Psi _k},\Psi _m\right) \left(
\Psi _m^{*},\int_0^\infty K_{\alpha \beta }^{(xy)}(\tau )e^{-iH_e(t)
\tau }D_\beta e^{iH_e(t)\tau }d\tau \Psi _k\right) +C.C.\right] \,, 
\]
with summation over all repeated indices and $k\neq m$ . Its structure
resembles that of $\Lambda _{test}$ and it also mixes 
kinetics of $\Psi $ and $\lambda $ .

What is for the case of sufficiently strong DS-bath coupling 
(intensive bath noise), one can make the substitution 
\[
\Psi (t)\Rightarrow S_x(t)\Psi (t)\,,
\,S_x(t)=\overleftarrow{\exp }
\left[ -i\int_0^tH(t^{\prime },x)dt^{\prime }\right] \,, 
\]
in Eqs.49-50, which allows to 
treat $(\Psi _m^{*},S_x^{+}(t)D_jS_x(t)\Psi _k) $ 
and similar quantities as fast noises and again 
apply Fokker-Plank ideology.

\subsection{Informational vacuum}

Let $W_{unif}(\Psi )$ be uniform distribution of $\Psi $ which is invariant
with respect to arbitrary rotations and thus independent on $\Psi $ at all.
Of course, this uniform measure is concentrated on the unit 
sphere, $(\Psi _m^{*},\Psi _k)$ $=\delta _{mk}$ , 
as well as $W(t,\lambda ,\Psi )$ in general, as Eq.50 does imply.

It is not hard to verify two facts. First, multiplication by $W_{unif}$
commutes with complete kinetic operator $\Lambda (t)$ and with any of its
four parts separately, 
\begin{equation}
\Lambda _sW_{unif}(\Psi )-W_{unif}(\Psi )\Lambda _s=0\,,\,\,
\,s\in \{test;\,free;\,xx;\,xy\}
\end{equation}
Second, kinetic equation (51) always has formal stationary solution 
\begin{equation}
W_{stat}(\lambda ,\Psi )=W_{unif}(\Psi )Y(\lambda )\,\,,\,\,\,
\Lambda (t)W_{stat}(\lambda ,\Psi )=0\,\,,
\end{equation}
where function $Y(\lambda )$ is common solution to the equations 
\begin{equation}
\left( \frac \partial {\partial \lambda _k}\lambda _k+
\sum_{m\neq k}\frac{\lambda _m+
\lambda _k}{\lambda _m-\lambda _k}\right) Y(\lambda )=0\,\,,
\,\,k=1..N\,\,
\end{equation}
The latter reads 
\begin{equation}
Y(\lambda )=\frac{\prod_{m<k}(\lambda _k-
\lambda _m)^2}{\prod_{j=1}^N\lambda _j^N}
\end{equation}
Moreover, each of the parts of $\Lambda (t)$ separately 
turns $W_{stat}(\lambda ,\Psi )$ into zero,
\[
\Lambda _sW_{stat}(\lambda ,\Psi )=0\,\,,\,\,
\,s\in \{test;\,free;\,xx;\,xy\} 
\]
The immovable point $W_{stat}(\lambda ,\Psi )$ can be named informational
vacuum since it says nothing about the system.

\subsection{Much more stationary states}

However, this stationary solution can be attracting point for solution of
Eq.51 at non-zero test parameter, $v(t)\neq 0$ , i.e. in absence of testing
only. To see this let us integrate both sides of Eq.51 over $\Psi $ . Since
the only part of $\Lambda (t)$ what has terms free of 
gradients $\ \partial /\partial \Psi _k$ 
and $\partial /\partial \Psi _k^{*}$ is $\Lambda _{test}$ , 
only its contribution survives after this integration: 
\begin{equation}
\frac \partial {\partial t}W_\lambda (t,\lambda )=-v(t)\sum_k\frac 
\partial {\partial \lambda _k}
\lambda _k\int (\Psi _k^{*},J\Psi _k)W(t,\lambda ,\Psi )d\Psi \,\,,
\end{equation}
\[
W_\lambda (t,\lambda )\equiv \int W(t,\lambda ,\Psi )d\Psi \,\, 
\]
Here $W_\lambda (t,\lambda )$ is marginal probability distribution of
eigenvalues. At $v(t)=0$ it does not change at all and remains 
atomic, $W_\lambda (t,\lambda )|_{v=0}$ $=\delta (\lambda -\lambda _0)$ , 
with $\lambda _0$ being initial values. Consequently, under a constant
perturbation, $u(t)=const$ , $W(t,\lambda ,\Psi )$ tends to a stationary
distribution,
\[
W_\infty (\lambda ,\Psi |\lambda _0)=\lim_{t\rightarrow 
\infty }\,W(t,\lambda ,\Psi )|_{v=0}\,\,, 
\]
which differs from (56) and remembers exact start eigenvalues. 
Hence, at $v(t)=0$ there is continuous infinite set of 
stationary solutions to Eq.51.

Interestingly, in general these degenerated solutions are not atomic with
respect to $\lambda $ , i.e. 
  $W_\infty (\lambda ,\Psi |\lambda _0)$ 
  $\neq \delta (\lambda -\lambda _0)$  
  $\cdot $ $\int W_\infty (\lambda ,\Psi |\lambda _0)d\Psi $ . 
This unusual property as well as the degeneracy itself
is reflection of ghost properties of $y(t)$ . The latters manifest
themselves in absence of second 
derivatives $\partial ^2/\partial \lambda ^2 $ in Eq.51, 
in spite of presence of joint derivatives 
  $\partial ^2/\partial \lambda \partial \Psi $ , 
  $\partial ^2/\partial \lambda \partial \Psi ^{*}$ . 
Therefore, strictly speaking, $\Lambda (t)$ is not purely elliptic
operator.

\subsection{Exclusion of eigenvectors}

In order to analyse the characteristic functional as presented by Eqs.47 or
48, we should consider marginal distribution $W_\lambda (t,\lambda )$ just
at $v(t)\neq 0$ , when it can change to arbitrary extent. Since then the
vacuum solution (56) is attractive, it is natural to use it as ground state
in the projection method. All the more, bath noise also 
pushes $W(t,\lambda ,\Psi )$ towards $W_{stat}(\lambda ,\Psi )$ 
making eigenvectors as much random and statistically 
independent on eigenvalues as possible. At the same
time, since the factor $Y(\lambda )$ is non-normalizable, this factored
solution never can be achieved explicitly. Always more or less correlations
between $\lambda $ and $\Psi $ do exist, and we must write
\[
W(t,\lambda ,\Psi )=W_\lambda (t,\lambda )W_{unif}(\Psi )+
\widetilde{W} (t,\lambda ,\Psi )\,\,,\,\,
\int \widetilde{W}(t,\lambda ,\Psi )d\Psi =0\,\,, 
\]
where $\widetilde{W}(t,\lambda ,\Psi )$ does not contribute 
to $W_\lambda (t,\lambda )$ .

Thus we merely introduce the projection operator: 
\[
\Pi W(\lambda ,\Psi )\equiv W_{unif}(\Psi )\int W(\lambda ,\Psi )d\Psi
\,,\,\,\,\widetilde{W}(t,\lambda ,\Psi )=(1-\Pi )W(t,\lambda ,\Psi ) 
\]
With using Eqs.55-56, Eq.51 yields 
\begin{equation}
\frac \partial {\partial t}\widetilde{W}=(1-\Pi )\Lambda (t)
\widetilde{W} +W_{unif}I(t)\,\,,\,\,
\,I(t)\equiv \left[ \Lambda (t)W_\lambda -
\int W_{unif}\Lambda (t)W_\lambda \,d\Psi \right]
\end{equation}
Here factor $I(t)$ in the source term is easily calculable: 
\begin{equation}
I(t)=-\sum_k\left( \frac \partial {\partial \lambda _k}
\lambda _k+\sum_{m\neq k}\frac{\lambda _m+\lambda _k}{\lambda _m-
\lambda _k}\right) W_\lambda (t,\lambda )[v(t)(
\Psi _k^{*},J\Psi _k)+(\Psi _k^{*},V(t)\Psi _k)]\,\,,
\end{equation}
with the traceless matrix $V(t)$ defined by the sum of commutators 
\begin{equation}
V(t)=i\sum_{\alpha \beta }\left[ 
\int_0^\infty K_{\alpha \beta }^{(xy)}(\tau )
\exp \{-iH_e(t)\tau \}D_\beta 
\exp \{iH_e(t)\tau \}d\tau \,,D_\alpha \right]
\end{equation}
Only $\Lambda _{test}$ and $\Lambda _{xy}$ contribute to the source giving
two similar terms in (61). Wee also took into account that 
\[
\int (\Psi _k^{*},M\Psi _k)W_{unif}(\Psi )d\Psi =(\,\text{Sp}\,M)/N 
\]
for any matrix $M$ . Because of this equality and of that the current matrix 
$J$ by its definition is traceless, only $\widetilde{W}$ contributes to
right-hand side in Eq.59.

\subsection{Kinetic equation for eigenvectors}

As usually in projection techniques, Eqs.59-62 ensure exclusion of a part of
variables, producing formally closed although non-local separate evolution
equation for eigenvalues. If introduce notations 
\[
A_{kq}\equiv (\Psi _k^{*},A\Psi _q)\,\,,\,\,\,
\left\langle (...)\right\rangle _{unif}\equiv 
\int (...)W_{unif}(\Psi )d\Psi \,\,, 
\]
\begin{equation}
\widehat{U}_v(t,t_0)=\overleftarrow{\exp }
\left\{ \int_{t_0}^t\left[ v(t^{\prime })(1-\Pi )
\Lambda _{test}+\Lambda _{free}(t^{\prime })+\Lambda _{xx}(t^{\prime })+
\Lambda _{xy}(t^{\prime })\right] dt^{\prime }\right\}\,\,,
\end{equation}
this equation reads 
\begin{equation}
\frac \partial {\partial t}W_\lambda (t,\lambda )=v(t)
\sum_{kq}\frac \partial {\partial \lambda _k}
\lambda _k\int_0^t\left\langle J_{kk}\widehat{U} _v(t,t^{\prime })
\left[ v(t^{\prime })J_{qq}+V_{qq}(t^{\prime })\right]
\right\rangle _{unif}\times
\end{equation}
\[
\times \left( \frac \partial {\partial \lambda _q}
\lambda _q+\sum_{m\neq q}
\frac{\lambda _m+\lambda _q}{\lambda _m-\lambda _q}
\right) W_\lambda (t^{\prime },\lambda )dt^{\prime }-v(t)
\sum_k\frac \partial {\partial
\lambda _k}\lambda _k\int J_{kk}\widehat{U}_v(t,0)
\widetilde{W}(0,\lambda ,\Psi )d\Psi 
\]
We take in mind Eqs.55-56 and the 
identities $(1-\Pi )\Lambda _s=\Lambda _s$ 
for $\,s\in $ $\{\,free;\,xx;\,xy\}$ .

The important point is that last term which represents 
a legacy of initial state,
\[
\widetilde{W}(0,\lambda ,\Psi )=\delta (\lambda -
\lambda _0)[\delta (\Psi -\Psi _0)-W_{unif}(\Psi )]\,\,, 
\]
may be insignificant even in spite of the degeneracy. 
The matter is that all the appointed stationary distributions 
do not depend on initial $\Psi $ distribution, 
therefore $\widehat{U}_v(t,0)$ $\widetilde{W}(0,\lambda ,\Psi ) $ 
must decay after a fixed finite time determined mainly by the diffusion
operator $\Lambda _{xx}$ , i.e. characteristic correlation time of
eigenvectors fluctuations. At the same time, as already Eq.59 shows, 
the evolution rate of $W_\lambda (t,\lambda )$ is dictated 
by magnitude of test function $v(t)$ . 
Since the latter serves eventually for differentiation 
by it, we may treat it as infinitesimally small parameter and 
hence $W_\lambda (t,\lambda )$ as slow varying, 
thus having all rights to neglect the last term.

\subsection{Kinetics of eigenvalues in hot bath}

Let the perturbation $u(t)$ which namely determines time dependence of
operators $\Lambda _{free}$ , $\Lambda _{xx}$ and $\Lambda _{xy}$ 
is not too strong and not too fast varying (for instance, $u(t)$ 
is constant). Then we can neglect time variations of $u(t)$ in the 
integrand kernel in Eq.64 while in (62) neglect $u(t)$ at all. 
Besides, let us neglect also the kernel
dependencies on both $v(t)$ and $\lambda $ , 
i.e throw away $\Lambda _{test}$ and $\Lambda _{xy}$ from the 
propagator (63). Formally, it is possible because the 
diffusion operator $\Lambda _{xx}$ itself quite ensures decay of
the kernel, while effects of operator $\Lambda _{xy}$ is partially 
took into account by the term containing $(\Psi _k^{*},V(t)\Psi _k)$ 
in the source (61). This term involves dissipation and represents the 
storage for non-zero dissipative mean current response.

Clearly, in such an approximation the current is modelled 
as Gaussian random process, and all effects of the degeneracy are asided. 
Neglecting $\Lambda _{xy}$ as compared 
with $\Lambda _{xx}$ corresponds to what can be named
''infinitely hot bath limit'', since according to Eq.23 action 
of $\Lambda _{xx}$ much overpowers action 
of $\Lambda _{xy}$ when bath temperature $\Theta $ grows to infinity. 
Thus we come to maximally simplified kinetic
equation for eigenvalues distribution, in the form 
\begin{equation}
\frac{\partial W_\lambda }{\partial t}=
\frac{\Gamma (t)}N\sum_k\frac \partial
{\partial \lambda _k}\lambda _k\left( \frac \partial {\partial 
\lambda _k}\lambda _k+\sum_{m\neq k}\frac{\lambda _m+
\lambda _k}{\lambda _m-\lambda _k}\right) W_\lambda
\end{equation}
Here the evolution rate $\Gamma (t)$ is determined by
\begin{equation}
\frac{\Gamma (t)}N=v(t)\int_0^\infty 
\left\langle (\Psi _0^{*},J\Psi _0)\exp
\{[\Lambda _{free}(t)+\Lambda _{xx}]\tau \}(\Psi _0^{*},V
\Psi _0)\right\rangle _{unif}d\tau +
\end{equation}
\[
+v(t)\int_0^\infty \left\langle (\Psi _0^{*},J\Psi _0)
\exp \{[\Lambda _{free}(t)+
\Lambda _{xx}]\tau \}(\Psi _0^{*},J\Psi _0)
\right\rangle _{unif}v(t-\tau )d\tau \,\,, 
\]
with $\Lambda _{free}(t)=$  $\Omega \{H_e\}-u(t)\Omega \{Q\}$ , 
vector $\Psi _0$ is any of columns of $\Psi $ , 
and matrix $V$ is given by (62) but under 
$H_e(t)$ replaced by $H_e$ . We assumed also that total number of
one-electron states (dimensionality of matrices) is large, $N>>1$ , 
which allows for many formal simplifications.

The first integral on right hand of Eq.66 describes mean current. We may
supppose that this integral turns into zero at $u(t)=0$ , because of a
proper symmetry of the matrix (62), i.e. that there is no spontaneous
current breaking time symmetry. We expect also at non-zero low-field
conductivity and sufficiently small and slow varying $u(t)$ (and besides
slow $v(t)$ ) expression (66) reduces to
\begin{equation}
\Gamma (t)\approx v(t)\frac \Delta \Theta u(t)+
\Delta v^2(t)=v(t)w(t)+\Delta v^2(t)\,\,,
\end{equation}
where $\Delta $ and $w(t)$ represent diffusivity and 
drift velocity of a separate electron.

Notice that Eq.65 closely resembles kinetic equations for transmission
matrix eigenvalues in the theory of statically disordered conductors [16].

\subsection{Current noise in Gaussian approximation}

Rather trivial model (65)-(67) is nevertheless useful probe to compare roles
of diffusive component of $W_\lambda (t,\lambda )$ evolution (described by
second-order derivatives $\partial ^2/\partial \lambda ^2$ in Eq.65), from
one hand, and of drift caused by mutual repulsion of eigenvalues (described
by ratios $(\lambda _m+\lambda _k)$ $/(\lambda _m-\lambda _k)$ ), from
another hand. Note that at possibly negative $\Gamma (t)$ values Eq.65
corresponds to mathematically incorrect time-reversed diffusion, but in fact
this implies nothing incorrect in final results of calculations.

Simple analysis shows that, at $N>>1$ , from the point of view of
statistical moments of $\lambda $ , all diffusive terms in Eq.65 occur
relatively insignificant. Therefore, in the asymptotical sense, for any
function $F(\lambda )$ we can write 
\begin{equation}
\frac d{dt}\left\langle F(\lambda )\right\rangle =
\Gamma (t)\left\langle \frac 1N\sum_k\left( 
\sum_{m\neq k}\frac{\lambda _k+\lambda _m}{\lambda _k-
\lambda _m}\right) \lambda _k\frac 
\partial {\partial \lambda _k} F(\lambda )\right\rangle \,\,,\,
\,\,\left\langle (...)\right\rangle \equiv
\int (...)W_\lambda d\lambda
\end{equation}
This circumstance is due to that many of 
ratios $(\lambda _m+\lambda _k)$ $/(\lambda _m-\lambda _k)$ 
are large numbers $\sim N$ , hence the drift
overpowers diffusion approximately $N$ times.

Particularly, in case (48), i.e. at fixed $N_{el} $  
(may be comparable with $N$ ), 
\begin{equation}
\sum_k\left( \sum_{m\neq k}
\frac{\lambda _k+\lambda _m}{\lambda _k-
\lambda _m }\right) \lambda _k\frac 
\partial {\partial \lambda _k}\sum^{\prime }
\lambda _{k_1}\lambda _{k_2}..
\lambda _{k_{N_{el}}}=N_{el}(N-N_{el})\sum^{\prime }
\lambda _{k_1}\lambda _{k_2}..\lambda _{k_{N_{el}}}
\end{equation}
(primed sum means that all indices are different one from another), and
Eq.68 exactly yields
\begin{equation}
\Xi (v)=\exp \left[ \theta N_{el}(N-N_{el})/N\right] \,\,,
\,\,\theta \equiv \int_0^t\Gamma (t^{\prime })dt^{\prime }
\end{equation}
This result corresponds to Gaussian current noise, with only those
electron-electron statistical correlations what come from Pauli principle
and Fermi statistics of electrons (or holes if $N_{el}\approx N$ ).

In case of grand canonical ensemble (47), we obtain 
\begin{equation}
\Xi (v)=\exp \left[ \theta \frac{N_{el}(N-N_{el})}N+
\frac{\theta ^2}2\left( 1-
\frac{2N_{el}}N\right) ^2\sum_kf_k(1-f_k)+...\right] \,\,,
\end{equation}
\[
\,f_k\equiv z\lambda _k/(1+z\lambda _k)\,\,,\,
\,N_{el}\equiv \sum_kf_k\,\,, 
\]
where $\lambda _k$ should be replaced by their initial values (i.e.
represent diagonal elements of $\rho (t=0)$ ). The points mean higher
degrees of $\theta $ . The second term in the exponent has no analogy in
(70) and formally corresponds to non-decaying flicker type current and
conductance fluctuations. But in fact this is nothing but artifact of 
grand canonical ensemble.

\subsection{Degeneracy and low-frequency noise}

Now let us be so courageous and discuss what will occur beyond 
the hot bath Gaussian approximation, i.e. 
if keep $v$ , $\lambda $ and $\partial /\partial \lambda $ 
dependencies of the kernel in Eq.64. Generally speaking,
we may expect that above mentioned strong degeneracy of stationary state
results in long-lasting statistical correlations when observing DS, i.e. in
low-frequency excess fluctuations of the current. Indeed, as Eq.59 shows,
measuring of the current transforms initial values $\lambda _0$ to one or
another different values $\lambda ^{\prime }$ which then remain for
arbitrary long time being unchangably encoded 
in $W_\infty (\lambda ,\Psi |\lambda ^{\prime })$ 
(or in similar time varying distribution if $u(t)$ is
not constant). Therefore, next measurement will be inevitably correlated
with previous one regardless of their time separation (this resembles
low-frequency Goldstonian excitations under breaking a continuous
degeneracy).

If this reasonings are true, then the differential operator kernel in Eq.64,
\begin{equation}
\widehat{R}_{kq}(t,t_0,v)\equiv v(t)\left\langle J_{kk}
\widehat{U} _v(t,t_0)\left[ v(t_0)J_{qq}+V_{qq}(t_0)\right] 
\right\rangle _{unif}\,\,,
\end{equation}
has slow decaying non-exponential tail closely related to continuous
degeneracy of ground state of the operator $(1-\Pi )\Lambda (t)$ which
generates semi-group $\widehat{U}_v(t,t_0)$ . Clearly, such a degeneracy
takes place (and besides, in opposite to $\Lambda (t)$ , even at 
arbitrary $v(t)$ ), hence, $\widehat{U}_v(t,t_0)$ is not strictly 
contracting semigroup. Directly, the degeneracy 
concerns $\lambda $ dependence of joint probability distribution, 
but indirectly it expands also to its $\Psi $
dependence, because of the $\lambda $ - $\Psi $ coupling produced by
operator $\Lambda _{xy}(t)$ . The hot bath Gaussian approximation 
considered neglects this coupling and therefore retains only 
rapidly decaying part (66) of the kernel (72) responsible for mean 
diffusivity and mean drift. The rest of the kernel must contain 
information on excess contribution to the current 
correlation function,
\[
K(t,t_0)\equiv \left\langle J(t)J(t_0)\right\rangle -
\left\langle J(t)\right\rangle \left\langle J(t_0)\right\rangle \,\,,
\]
including conductance fluctuations, and besides on higher-order cumulants of
current. The former, in its turn, consists of two parts whose physical sense
is as follows.

When the bath noise $x(t)$ transforms coherent electron motion into
diffusion (plus drift as characteristics of its asymmetry), diffusivity and
mobility of any electron depend on random time-varying non-uniformities of
density of other electrons, i.e. on random occupancies of one-electron
states. As it will be demonstrated elsewhere, this fact results in flicker
type conductance fluctuations with spectrum $\propto f^{-1/2}$ , at
frequencies higher then inverse time of electron diffusion through overall
system. This excess noise arises merely from Pauli principle and has no
relation to $\lambda $ - $\Psi $ coupling. The only role 
of $\Lambda _{xy}(t) $ in this mechanism is establishing more or less 
statistical hierarchy (non-equiparticity) of electron energies, 
in contrary to $x(t)$ pumping, and thus translate diffusivity 
fluctuations into drift ones.

Other mechanism of excess low-frequency noise is due, in principle, just to
the $\lambda $ - $\Psi $ coupling. As we already pointed out, the reverse DS
action onto the bath, representing dissipation, makes the latter mediator of
indirect electron-electron interaction through bath. This is in rich analogy
with action of the $x^{\prime }(t)$ - $y^{\prime }(t)$ noise in Eqs.28-29
which is equivalent to direct electron-electron interaction. Hence, to keep
operator $\Lambda _{xy}(t)$ in (72) means to take into account such an
interaction. But our experience gave evidents [5,8] that particle-particle
interaction itself is sufficient natural cause of a true flicker noise, i.e.
noise without low-frequency spectrum saturation (1/f fluctuations of
self-diffusivity and mobility of particles). In [1,2,4,5] it was explained
why this noise can be insensible to spatial and temporal scales of a system.

\subsection{Current correlation function}

To connect $K(t,t_0)$ and the kernel (72), let us suppose, for simplicity
and visuality, that external perturbation $u(t)$ is time-independent and
that after not a long time a mean value of the 
current, $\overline{J} =$  $\left\langle J(t)\right\rangle $ , 
tends to a constant, while its correlation function becomes 
depending on time difference only, $K(t-t_0)$ .
Then we can put on the test function $v(t)$ also constant, $v=$ $const$ ,
thus testing integrated current, i.e. transported 
charge, $\int_0^tJ(t^{\prime })dt^{\prime }$ . 
This simplifications bring comforts of Laplas transformation. 

Applying the latter to the characteristic functional, 
at sufficiently small $v$ and $p$ we have 
\begin{equation}
\widetilde{\Xi }(v)\equiv \int_0^\infty 
\Xi (v)e^{-pt}dt=[p-v\overline{J}-v^2
\widetilde{K}(p)-...]^{-1}\,\,,\,\,
\,\widetilde{K}(p)\equiv 
\int_0^\infty K(\tau )e^{-p\tau }d\tau \,,
\end{equation}
where the dots replace higher-order terms of $v$ expansion. From the other
hand, the same quantity can be expressed from Eqs.47-48 and evolution
equation (64) (throwing out last term of (64)), 
\begin{equation}
\widetilde{\Xi }(v)=F^{-1}(\lambda )
\left[ p-v\widehat{G}_1(p)-v^2\widehat{G} _2(p)-...
\right] ^{-1}F(\lambda )
\end{equation}
Here differential operators $\widehat{G}_n(p)$ acting on $\lambda $ are
defined by $v$ expansion of kernel (72), and $\lambda $ stand for the
initial eigenvalues. Direct formal expansion gives 
\[
\widehat{G}_1(p)=\sum_{kq}[\partial _q+
\phi _q(\lambda )]\left\langle J_{kk}[p-
\Lambda _0^T]^{-1}V_{qq}\right\rangle _{unif}\partial _k\,\,,
\]
\begin{equation}
\widehat{G}_2(p)=\sum_{kq}[\partial _q+\phi _q(\lambda )]
\left\langle J_{kk}[p-\Lambda _0^T]^{-1}J_{qq}
\right\rangle _{unif}\partial _k+
\end{equation}
\[
+\sum_{kq}[\partial _q+\phi _q(\lambda )]
\left\langle J_{kk}[p-\Lambda _0^T]^{-1}(1-\Pi )
\Lambda _{test}^T[p-\Lambda _0^T]^{-1}V_{qq}
\right\rangle _{unif}\partial _k\,\,,
\]
where the shortened notations are used 
\[
\partial _k\equiv \lambda _k\frac 
\partial {\partial \lambda _k}\,\,,\,\,\,
\phi _k(\lambda )\equiv \sum_{m\neq k}
\frac{\lambda _k+\lambda _m }{\lambda _k-\lambda _m}\,\,,\,\,
\,\Lambda _0\equiv \Lambda _{free}+\Lambda _{xx}+\Lambda _{xy}\,
\]
The superscript ''$T$'' is symbol of conjugation 
with respect to $\lambda $ in the Sturm-Liouville sense, 
that is reversing of order of all
differentiations $\partial /\partial \lambda $ and multiplications 
by any function of $\lambda $ and 
changing $\partial /\partial \lambda $ 
by $-\partial /\partial \lambda $ . 

Comparing linear and quadratic terms in 
formulas (73) and (74) we see that
\begin{equation}
\overline{J}=F^{-1}(\lambda )\widehat{G}_1(p)F(\lambda )\,\,,
\end{equation}
\[
\widetilde{K}(p)=F^{-1}(\lambda )
\widehat{G}_2(p)F(\lambda )+p^{-1}\{F^{-1}(\lambda )
\widehat{G}_1^2(p)F(\lambda )-[F^{-1}(\lambda ) 
\widehat{G}_1(p)F(\lambda )]^2\}
\]
Thus, mean current really tends to a constant 
if $\widehat{G}_1(p\rightarrow 0)$ has finite limit. 
Besides, the natural possibility is if in this 
long-time limit $F(\lambda ) $ is eigenfunction of 
operator  $\widehat{G}_1(p\rightarrow 0)$ . Then Laplas
transform of the current correlator reduces to
\begin{equation}
\widetilde{K}(p)\approx F^{-1}(\lambda )\widehat{G}_2(p)F(\lambda )
\end{equation}
Hence,  $F(\lambda )$ may be also eigenfunction 
of  $\widehat{G}_2(p\rightarrow 0)$ . Notice that operators (75) 
always can have polynomial eigenfunctions like presented by (48) 
since these operators (as well as complete 
operators $\,\Lambda _0$ , $\,\Lambda _{test}$ ) are
insensible to arbitrary scaling of $\lambda $ . Of course, 
both $\overline{J} $ and excess part 
of  $\widetilde{K}(p)$ are results of external
perturbation, $u$ , hidden  
in  $\Lambda _{free}=$  $\Omega \{H_e\}-u\Omega \{Q\}$ . 

\subsection{Resume}

For the meanwhile, our main results are, first, reduction of many-electron
problems to investigation of stochastic one-electron density 
matrix, $\rho (t)$ , which obeys exact Langevin equation (20), 
and, second, formulation of approximate Fokker-Plank kinetics for 
eigenvalues and eigenvectors of this matrix.

Equivalently, one could consider statistical moments 
  $\left\langle \rho(t)\otimes ...\otimes \rho (t)\right\rangle _b$ , 
i.e. mean strict products of many copies of $\rho (t)$ , and 
derive from (20) evolution equations for them or, instead, 
express solution to (20) and then the moments in terms of 
many-path integrals with counting intersections (raprochements) of 
paths (like it was considered in [14] with respect to single copy).
Under the weak short-correlated bath noise approximation, evolution 
equations are time-local and then, at least, right-hand side of 
equation for  $\left\langle \rho(t) \right\rangle _b$ 
can be represented in standard Lindblad form for generators 
of quantum contracting semigroups (although in our case 
this is not natural form). But the Langevin equation (20), 
which in our case realizes stochastic dilation of a 
dissipative semigroup behaviour, lies rather far from Langevin 
equations under use in modern ''quantum stochastic calculus'' [13] or 
in well developed bosonic stochastic calculus [15]. The peculiarity of 
Eq.20 is that it takes into account phase volume exchange between 
dynamical subsystem (electrons) and thermal bath and, therefore, 
the dilation it gives is not unitary in momentary sense although 
eventually (after complete averaging) works as unitary one. To the 
best of our knowledge, such sort of Langevin equations was still 
unknown. If it is so then Eq.20 itself also can be qualified as 
result of this paper. 

In the frame of this equation, we came to very intriguing task about
spectral and other properties of the Fokker-Planck operator 
  $\Lambda _0=$ $\Lambda _{free}+$ $\Lambda _{xx}+$ $\Lambda _{xy}$ 
whose peculiarities are non-self-adjointness, non-ellipticity 
(induced by that of $\Lambda _{xy}$ ) and continuous infinitness 
of Ker $\Lambda _0$ . Particularly, Fokker-Planck approximation 
turns into exact theory if throw out thermal bath at all but 
instead include delta-correlated $x^{\prime }(t)$ - $y^{\prime }(t)$ 
noise equivalent to pair electron-electron interactions. In
this case $\Lambda _{xx}=0$ and ''bad'' operator $\Lambda _{xy}$ 
becomes responsible for everything (perhaply, in particular, 
for such ''bad'' phenomenon as flicker noise).

\,\,

I acknowledge Dr. I.Krasnyuk and Dr. Yu.Medvedev for 
useful conversations. 

\,\,

\subsection{References}        

1. G.N.Bochkov and Yu.E.Kuzovlev, Sov.Phys.-Usp., 26, 829 (1983).

2. Yu.E.Kuzovlev and G.N.Bochkov, Izv. VUZov. -Radiofizika, 26, 310 (1983),
transl. in Radiophysics and Quantum Electronics, No. 3, 1983.

3. G.N.Bochkov and Yu.E.Kuzovlev, Izv. VUZov. -Radiofizika, 27, 1151 (1984),
transl. in Radiophysics and Quantum Electronics, No. 9, 1984.

4. G.N.Bochkov and Yu.E.Kuzovlev. On the theory of 1/f-noise. Preprint No.
195. NIRFI, Gorkii, USSR, 1985 (in Russian).

5. Yu.E.Kuzovlev, Sov.Phys.-JETP, 67, No. 12, 2469 (1988).

6. Yu.E.Kuzovlev, JETP, 84, No. 6, 1138 (1997).

7. Yu.E.Kuzovlev, Phys.Lett., A 194, 285 (1994).

8. Yu.E.Kuzovlev, cond-mat/9903350.

9. Yu.E.Kuzovlev, cond-mat/0004398.

10. Yu.E.Kuzovlev, Yu.V.Medvedev and A.M.Grishin, cond-mat/0010447.

11. Yu.E.Kuzovlev, Yu.V.Medvedev and A.M.Grishin, JETP Lett., 72, 574 (2000).

12. N.S.Krylov. Works on the foundations of statistical physics. Princeton,
1979.

13. A.S.Holevo. Quantum stochastic calculus. In ''Modern problems of
mathematics. Novel adventures. V.36. Editor R.V.Gamkrelidze''. Moscow,
VINITI, 1991 (in Russian).

14. R.P.Feynman and A.R.Hibbs. Quantum mechanics and path integrals. N.-Y.,
1965.

15. M.Lax. Fluctuation anf coherence phenomena in classical and quantum
physics. N.-Y., 1968.

16. C.W.J.Beenakker, Rev.Mod.Phys., 69, No. 3, 731 (1997).

\end{document}